\documentclass[12pt]{article}
\usepackage{amssymb,amsmath,epsfig}
\begin{document}
\title{\bf Non-Vacuum Bianchi Types $I$ and $V$ in $f(R)$ Gravity}

\author{M. Sharif \thanks{msharif@math.pu.edu.pk} and M. Farasat
Shamir \thanks{frasat@hotmail.com}\\\\
Department of Mathematics, University of the Punjab,\\
Quaid-e-Azam Campus, Lahore-54590, Pakistan.}

\date{}

\maketitle
\begin{abstract}
In a recent paper \cite{1}, we have studied the vacuum solutions of
Bianchi types $I$ and $V$ spacetimes in the framework of metric
$f(R)$ gravity. Here we extend this work to perfect fluid solutions.
For this purpose, we take stiff matter to find energy density and
pressure of the universe. In particular, we find two exact solutions
in each case which correspond to two models of the universe. The
first solution gives a singular model while the second solution
provides a non-singular model. The physical behavior of these models
has been discussed using some physical quantities. Also, the
function of the Ricci scalar is evaluated.
\end{abstract}

\section{Introduction}

The most interesting phenomenon, which the modern day physics
deals with today, is the accelerated expansion of the universe.
The explanation for this universe expansion has both theoretical
as well as experimental background \cite{2}. It has been found
that most of the universe contains dark energy and dark matter.
The cosmological constant is shown \cite{3} to be an alternative
for dark energy. $f(R)$ theory of gravity is an attractive
candidate as an alternative theory of gravity in which a general
function of the Ricci scalar, $f(R)$, replaces $R$ in the standard
Einstein-Hilbert Lagrangian. This provides \cite{4} an easy
unification of early time inflation and late time acceleration.
The theory also gives a natural gravitational alternative to dark
energy.

Lobo \cite{05} gave a brief review of some of the modified
theories of gravity that address dark energy and the dark matter
problems. In another review \cite{005}, Faraoni discussed the
explanation of the cosmic acceleration alternative to dark energy
in the various versions of $f(R)$ theories of gravity. He
investigated the successes of $f(R)$ gravity together with the
challenges imposed by minimal criteria for their viability.
Sotiriou and Faraoni \cite{0005} presented some important aspects
of $f(R)$ theories of gravity in Metric, Palatini and
Metric-Affine formalisms. They discussed the motivation, actions,
field equations, equivalence with other theories, cosmological
aspects and constraints, viability criteria and astrophysical
applications.

The $f(R)$ theory of gravity is one of the modified theories which
is considered most suitable due to cosmologically important $f(R)$
models. These models consist of higher order curvature invariants
as functions of the Ricci scalar. Viable $f(R)$ gravity models
\cite{5} have been proposed which show the unification of
early-time inflation and late-time acceleration. It is hoped that
the problem of dark matter can be addressed by using viable $f(R)$
gravity models. Multam$\ddot{a}$ki and Vilja \cite{6} investigated
spherically symmetric vacuum solutions in $f(R)$ theory. The same
authors \cite{7} also studied the perfect fluid solutions and
showed that pressure and density did not uniquely determine
$f(R)$. Cognola et al. \cite{8} investigated $f(R)$ gravity at
one-loop level in de-Sitter universe. Capozziello et al. \cite{9}
explored spherically symmetric solutions of $f(R)$ theories of
gravity via the Noether symmetry approach. Recently, Sharif and
Kausar \cite{010} studied non-vacuum static spherically symmetric
solutions in $f(R)$ gravity.

Hollenstein and Lobo \cite{10} analyzed exact solutions of static
spherically symmetric spacetimes in $f(R)$ gravity coupled to
non-linear electrodynamics. Azadi et al. \cite{11} studied
cylindrically symmetric vacuum solutions in this theory. Momeni and
Gholizade \cite{12} extended cylindrically symmetric solutions in a
more general way. We have explored static plane symmetric vacuum
solutions \cite{13} in $f(R)$ gravity. The field equations are
solved using the assumption of constant scalar curvature which may
be zero or non-zero. However, very few attempts have been made so
far for solutions with non-constant scalar curvature.

Bianchi types $I$ and $V$ spacetimes play an important role in the
study of universe. Lorenz-Petzold \cite{0018} studied exact
Brans-Dicke Bianchi type $I$ solutions with a cosmological constant.
The same author \cite{0019} also derived solutions for locally
rotationally symmetric (LRS) Bianchi type $V$ spacetime in the
Brans-Dicke theory of gravitation. It was found that solutions
represented anisotropic cosmological models filled with stiff matter
and an electromagnetic null field. Berman \cite{019} introduced a
different method to solve the field equations by using the variation
law of Hubble parameter. The main feature of the variation law is
that it gives a constant value of the deceleration parameter. Kumar
and Singh \cite{022} investigated perfect fluid solutions using
Bianchi type $I$ spacetime in scalar-tensor theory. Singh
\cite{0220} investigated LRS Bianchi type $V$ cosmology with heat
flow in Scalar-Tensor theory. Paul et al. \cite{0022} investigated
FRW cosmologies in $f(R)$ gravity. In a recent paper \cite{1}, we
have studied the vacuum solutions of Bianchi types $I$ and $V$
spacetimes in the framework of $f(R)$ gravity.

In this paper, we focuss our attention to explore the perfect
fluid solutions of Bianchi types $I$ and $V$ spacetimes in metric
$f(R)$ gravity. For this purpose, we take the case of stiff matter
to find energy density and pressure of the universe. The paper is
organized as follows: In section \textbf{2}, we briefly give the
field equations in metric $f(R)$ gravity. Sections \textbf{3} and
\textbf{4} are used to find some exact solutions of Bianchi types
$I$ and $V$ spacetimes respectively. In the last section, we
discuss the results.

\section{$f(R)$ Gravity Formalism}

The action for $f(R)$ gravity is given by
\begin{equation}\label{1}
S=\int\sqrt{-g}(\frac{1}{16\pi{G}}f(R)+L_{m})d^4x,
\end{equation}
where $f(R)$ is a general function of the Ricci scalar and $L_{m}$
is the matter Lagrangian. The corresponding field equations are the
following:
\begin{equation}\label{2}
F(R)R_{\mu\nu}-\frac{1}{2}f(R)g_{\mu\nu}-\nabla_{\mu}
\nabla_{\nu}F(R)+g_{\mu\nu}\Box F(R)=\kappa T_{\mu\nu},
\end{equation}
where $F(R)\equiv
df(R)/dR,~\Box\equiv\nabla^{\mu}\nabla_{\mu},~\nabla_{\mu}$ is the
covariant derivative and $T_{\mu\nu}$ is the standard matter
energy-momentum tensor derived from the Lagrangian $L_m$. When we
contract the field equations, it follows that
\begin{eqnarray}\label{9}
F(R)R-2f(R)+3\Box F(R)=\kappa T.
\end{eqnarray}
Using this equation in Eq.(\ref{2}), the field equations take the
form
\begin{equation}\label{10}
F(R)R_{\mu\nu}-\nabla_{\mu}\nabla_{\nu}F(R)-\kappa T_{\mu\nu}
=[\frac{F(R)R-\Box F(R)-\kappa T}{4}]g_{\mu\nu}.
\end{equation}
Thus we have eliminated $f(R)$ from the field equations which
helps us to solve the field equations.

\section{Exact Bianchi Type $I$ Solutions}

In this section we find exact solutions of Bianchi I spacetime in $f(R)$
gravity and some physical quantities.

\subsection{Field Equations and Some Physical Quantities}

The line element of Bianchi type $I$ spacetime is given by
\begin{equation}\label{6}
ds^{2}=dt^2-A^2(t)dx^2-B^2(t)dy^2-C^2(t)dz^2,
\end{equation}
where $A,~B$ and $C$ are cosmic scale factors. The corresponding
Ricci scalar is
\begin{equation}\label{7}
R=-2[\frac{\ddot{A}}{A}+\frac{\ddot{B}}{B}+\frac{\ddot{C}}{C}
+\frac{\dot{A}\dot{B}}{AB}+\frac{\dot{B}\dot{C}}{BC}+\frac{\dot{C}\dot{A}}{CA}],
\end{equation}
where dot means derivative with respect to $t$. The energy-momentum
tensor for perfect fluid yields
\begin{equation}\label{04}
T_{\mu\nu}=(\rho + p)u_\mu u_\nu-pg_{\mu\nu},
\end{equation}
satisfying the equation of state
\begin{equation}\label{004}
p=w\rho,\quad0\leq w\leq1,
\end{equation}
where $\rho$ and $p$ are energy density and pressure of the fluid
while $u_\mu=\sqrt{g_{00}}(1,0,0,0)$ is the four-velocity in
co-moving coordinates. Since the metric (\ref{6}) depends only on
$t$, Eq.(\ref{10}) yields a set of differential equations for
$F(t)$, $A,~B,~C,~\rho$ and $p$. Thus the subtraction of the
$00$-component and $11$-component gives
\begin{equation}\label{12}
-\frac{\ddot{B}}{B}-\frac{\ddot{C}}{C}
+\frac{\dot{A}\dot{B}}{AB}+\frac{\dot{C}\dot{A}}{CA}
+\frac{\dot{A}\dot{F}}{AF}-\frac{\ddot{F}}{F}-\frac{\kappa}{F}(\rho+p)=0.
\end{equation}
Similarly, we get two more independent equations
\begin{eqnarray} \label{13}
-\frac{\ddot{A}}{A}-\frac{\ddot{C}}{C}
+\frac{\dot{A}\dot{B}}{AB}+\frac{\dot{B}\dot{C}}{BC}
+\frac{\dot{B}\dot{F}}{BF}-\frac{\ddot{F}}{F}-\frac{\kappa}{F}(\rho+p)=0,\\\label{14}
-\frac{\ddot{A}}{A}-\frac{\ddot{B}}{B}
+\frac{\dot{B}\dot{C}}{BC}+\frac{\dot{C}\dot{A}}{CA}
+\frac{\dot{C}\dot{F}}{CF}-\frac{\ddot{F}}{F}-\frac{\kappa}{F}(\rho+p)=0.
\end{eqnarray}
The conservation equation, $T^{\mu\nu}_{~~;\nu}=0$, leads to
\begin{equation}\label{0142}
\dot{\rho}+(p+\rho)[\frac{\dot{A}}{A}+\frac{\dot{B}}{B}+\frac{\dot{C}}{C}]=0.
\end{equation}
Finally, we have four differential equations with six unknowns
namely $A,~B$, $C,~F,~\rho$ and $p$. The solution of these equations
is discussed in the next subsection. In the following we give
definition of some physical quantities.

We define the average scale factor and the volume scale factor
respectively
\begin{eqnarray}\label{8}
a=\sqrt[3]{ABC},\quad V=a^3=ABC.
\end{eqnarray}
The generalized mean Hubble parameter $H$ is given in the form
\begin{equation}\label{008}
H=\frac{1}{3}(H_1+H_2+H_3),
\end{equation}
where
$H_1=\frac{\dot{A}}{A},~H_2=\frac{\dot{B}}{B},~H_3=\frac{\dot{C}}{C}$
are the directional Hubble parameters in the directions of $x,~y$
and $z$ axis respectively. Using Eqs.(\ref{8}) and (\ref{008}), we
obtain
\begin{equation}\label{0008}
H=\frac{1}{3}\frac{\dot{V}}{V}=\frac{1}{3}(H_1+H_2+H_3)=\frac{\dot{a}}{a}.
\end{equation}
The mean anisotropy parameter $A$ is given by
\begin{equation}\label{0000009}
A=\frac{1}{3}\sum^3_{i=1}(\frac{H_i-H}{H})^2.
\end{equation}
The expansion scalar $\theta$ and shear
scalar $\sigma^2$ are defined as follows
\begin{eqnarray}\label{09}
\theta&=&u^\mu_{;\mu}=\frac{\dot{A}}{A}+\frac{\dot{B}}{B}+\frac{\dot{C}}{C},\\
\label{00009}
\sigma^2&=&\frac{1}{2}\sigma_{\mu\nu}\sigma^{\mu\nu}
=\frac{1}{3}[(\frac{\dot{A}}{A})^2+(\frac{\dot{B}}{B})^2
+(\frac{\dot{C}}{C})^2-\frac{\dot{A}\dot{B}}{AB}-\frac{\dot{B}\dot{C}}{BC}
-\frac{\dot{C}\dot{A}}{CA}],
\end{eqnarray}
where
\begin{equation}\label{009}
\sigma_{\mu\nu}=\frac{1}{2}(u_{\mu;\alpha}h^\alpha_\nu+u_{\nu;\alpha}h^\alpha_\mu)
-\frac{1}{3}\theta h_{\mu\nu},
\end{equation}
$h_{\mu\nu}=g_{\mu\nu}-u_{\mu}u_{\nu}$ is the projection tensor.
In thermodynamics, the entropy of the universe is
given by
\begin{equation}\label{901}
TdS=d(\rho V)+pdV.
\end{equation}

\subsection{Solution of the Field Equations}

Subtracting Eqs.(\ref{13}), (\ref{14}) and (\ref{14}) from
Eqs.(\ref{12}), (\ref{13}) and (\ref{12}), we get respectively
\begin{eqnarray}\label{015}
\frac{\ddot{A}}{A}-\frac{\ddot{B}}{B}
+\frac{\dot{C}}{C}(\frac{\dot{A}}{A}-\frac{\dot{B}}{B})
+\frac{\dot{F}}{F}(\frac{\dot{A}}{A}-\frac{\dot{B}}{B})=0,\\\label{016}
\frac{\ddot{B}}{B}-\frac{\ddot{C}}{C}
+\frac{\dot{A}}{A}(\frac{\dot{B}}{B}-\frac{\dot{C}}{C})
+\frac{\dot{F}}{F}(\frac{\dot{B}}{B}-\frac{\dot{C}}{C})=0,\\\label{017}
\frac{\ddot{A}}{A}-\frac{\ddot{C}}{C}
+\frac{\dot{B}}{B}(\frac{\dot{A}}{A}-\frac{\dot{C}}{C})
+\frac{\dot{F}}{F}(\frac{\dot{A}}{A}-\frac{\dot{C}}{C})=0.
\end{eqnarray}
These equations are exactly the same as given by Eqs.(18)-(20) in \cite{1}.
Thus we can write the metric functions
explicitly as
\begin{eqnarray}\label{19}
A=ap_1\exp[{q_1\int\frac{dt}{a^3F}}],\\\label{20}
B=ap_2\exp[{q_2\int\frac{dt}{a^3F}}],\\\label{21}
C=ap_3\exp[{q_3\int\frac{dt}{a^3F}}],
\end{eqnarray}
where
\begin{equation}\label{22}
p_1=({d_1}^{-2}{d_2}^{-1})^{\frac{1}{3}},\quad
p_2=(d_1{d_2}^{-1})^{\frac{1}{3}},\quad
p_3=(d_1{d_2}^2)^{\frac{1}{3}},\quad p_1p_2p_3=1
\end{equation}
and
\begin{equation}\label{23}
q_1=-\frac{2c_1+c_2}{3},\quad q_2=\frac{c_1-c_2}{3},\quad
q_3=\frac{c_1+2c_2}{3},\quad q_1+q_2+q_3=0,
\end{equation}
$c_i$ and $d_i$ are constants of integration. Using power law
relation between $F$ and $a$ \cite{1}, we have
\begin{equation}\label{25}
F=ka^m,
\end{equation}
where $k$ is the constant of proportionality, $m$ is any integer
(here taken as $2$) and $a$ is given by
\begin{eqnarray}\label{30}
a&=&(nlt+k_1)^{\frac{1}{n}},\quad n\neq0\nonumber\\
a&=&k_2\exp(lt),\quad n=0,
\end{eqnarray}
$k_1$ and $k_2$ are constants of integration. It is mentioned here
that we have used $H=la^{-n},~l>0,~n\geq0$ to get the above
equation. Thus we obtain two values of the average scale factor
corresponding to two different models of the universe.

\subsection{Model of the Universe when $n\neq0$}

For this model, $F$ becomes $F=k(nlt+k_1)^{-\frac{2}{n}}$ and the
corresponding metric coefficients $A,~B$ and $C$ turn out to be
\begin{eqnarray}\label{35}
A&=&p_1(nlt+k_1)^{\frac{1}{n}}\exp[\frac{q_1(nlt+k_1)^
{\frac{n-1}{n}}}{kl(n-1)}],\quad n\neq1\\\label{36}
B&=&p_2(nlt+k_1)^{\frac{1}{n}}\exp[\frac{q_2(nlt+k_1)^
{\frac{n-1}{n}}}{kl(n-1)}],\quad n\neq1\\\label{37}
C&=&p_3(nlt+k_1)^{\frac{1}{n}}\exp[\frac{q_3(nlt+k_1)^
{\frac{n-1}{n}}}{kl(n-1)}],\quad n\neq1.
\end{eqnarray}
The mean generalized Hubble parameter and the volume scale factor become
\begin{equation}\label{39}
H=\frac{l}{nlt+k_1},\quad V=(nlt+k_1)^\frac{3}{n}.
\end{equation}
The mean anisotropy parameter $A$ turns out to be
\begin{equation}\label{0040}
A=\frac{{q_1}^2+{q_2}^2+{q_3}^2}{3l^2k^2(nlt+k_1)^{\frac{2}{n}-2}}.
\end{equation}
The expansion $\theta$ and shear scalar
$\sigma^2$ are given by
\begin{equation}\label{040}
\theta=\frac{3l}{nlt+k_1},\quad
\sigma^2=\frac{{q_1}^2+{q_2}^2+{q_3}^2}{2k^2(nlt+k_1)^{\frac{2}{n}}}.
\end{equation}

For stiff matter ($w=1$), we have $p=\rho$. Thus the energy density
and pressure of the universe become
\begin{equation}\label{0041}
2\kappa p=2\kappa\rho=-\frac{6kl^2}{(nlt+k_1)^{\frac{2}{n}+2}}
-\frac{{q_1}^2+{q_2}^2+{q_3}^2}{k(nlt+k_1)^{\frac{4}{n}}}.
\end{equation}
The entropy of universe is given by
\begin{equation}\label{902}
TdS=\frac{1}{\kappa}[6kl^3(n-2)(nlt+k_1)^{\frac{1}{n}-3}-
\frac{l}{k}({q_1}^2+{q_2}^2+{q_3}^2)(nlt+k_1)^{-\frac{1}{n}-1}].
\end{equation}
Also, Eq.(\ref{0142}) leads to
\begin{equation}\label{0144}
\rho=\frac{c}{V^2},
\end{equation}
where c is an integration constant. It is mentioned here that this
value of $\rho$, when compared with the value obtained in
Eq.(\ref{0041}), gives a constraint
\begin{equation}\label{0145}
\kappa c+3kl^2=0
\end{equation}
which holds only when $n=2$ and ${q_1}^2+{q_2}^2+{q_3}^2=0$. The
function of Ricci scalar, $f(R)$ is
\begin{equation}\label{41}
f(R)=\frac{k}{2}(nlt+k_1)^\frac{-2}{n}R+3kl^2(n-2)(nlt+k_1)^\frac{-2n-2}{n},
\end{equation}
where $R\equiv R_1=6l^2(n-2)(nlt+k_1)^{-2}$. For a special case
$n=\frac{1}{2}$, $f(R)$ turns out to be
\begin{equation}\label{0321}
f(R)=\frac{k}{2}(-\frac{R}{9l^2})^{\frac{1}{n}}
-\frac{9kl^2}{2}(-\frac{R}{9l^2})^{\frac{1}{n}+1}
\end{equation}
which gives $f(R)$ in terms of $R$.

\subsection{Model of the Universe when $n=0$}

Here the metric coefficients take the form
\begin{eqnarray}\label{43}
A&=&p_1k_2\exp(lt)\exp[-\frac{q_1\exp(-lt)}{klk_2}],\\\label{44}
B&=&p_2k_2\exp(lt)\exp[-\frac{q_2\exp(-lt)}{klk_2}],\\\label{37}
C&=&p_3k_2\exp(lt)\exp[-\frac{q_3\exp(-lt)}{klk_2}].
\end{eqnarray}
The directional Hubble parameters $H_i$ and the mean generalized
Hubble parameter will become
\begin{equation}\label{44}
H_i=l+\frac{q_i}{kk_2}\exp(-lt),\quad H=l
\end{equation}
while the volume scale factor turns out to be
\begin{equation}\label{46}
V={k_2}^3\exp(3lt).
\end{equation}
The mean anisotropy parameter $A$ becomes
\begin{equation}\label{0046}
A=[\frac{{q_1}^2+{q_2}^2+{q_3}^2}{3l^2k^2{k_2}^2}]\exp(-2lt)
\end{equation}
while the quantities $\theta$ and $\sigma^2$ are given by
\begin{equation}\label{046}
\theta=3l,\quad
\sigma^2=[\frac{{q_1}^2+{q_2}^2+{q_3}^2}{2k^2{k_2}^2}]\exp(-2lt).
\end{equation}

For stiff matter, the energy density and pressure turn out to be
\begin{equation}\label{00046}
2\kappa p=2\kappa\rho=-\frac{6kl^2\exp(-2lt)}{{k_2}^2}
-\frac{({q_1}^2+{q_2}^2+{q_3}^2)\exp(-4lt)}{k{k_2}^4}.
\end{equation}
The corresponding entropy is
\begin{equation}\label{904}
TdS=\frac{1}{\kappa}[-12kk_2l^3\exp(lt)-
\frac{l}{kk_2}({q_1}^2+{q_2}^2+{q_3}^2)\exp(-lt)].
\end{equation}
The constraint equation with the condition,
${q_1}^2+{q_2}^2+{q_3}^2=0$, is given by
\begin{equation}\label{0146}
\kappa c-6k{k_2}^4l^2\exp(4lt)=0.
\end{equation}
The function of Ricci scalar, $f(R)$, takes the form
\begin{equation}\label{47}
f(R)=\frac{k}{{2k_2}^2}\exp(-2lt)(R-12l^2)
\end{equation}
which reduces to
\begin{equation}\label{047}
f(R)=\sqrt{\frac{3k^3l^2}{2\kappa c}}[R-12l^2]
\end{equation}
using the constraint Eq.(\ref{0146}). This corresponds to the general
function $f(R)$ \cite{16},
\begin{equation}\label{0047}
f(R)=\sum a_n R^n,
\end{equation}
where $n$ may take the values from negative or positive.

\section{Exact Bianchi Type $V$ Solutions}

Here we shall find exact solutions of the Bianchi type $V$ spacetime.

\subsection{Field Equations}

The metric for the Bianchi type $V$ spacetime is
\begin{equation}\label{56}
ds^{2}=dt^2-A^2(t)dx^2-e^{2mx}[B^2(t)dy^2+C^2(t)dz^2],
\end{equation}
where $A,~B$ and $C$ are cosmic scale factors and $m$ is an
arbitrary constant. The corresponding Ricci scalar is
\begin{equation}\label{57}
R=-2[\frac{\ddot{A}}{A}+\frac{\ddot{B}}{B}+\frac{\ddot{C}}{C}-\frac{3m^2}{A^2}
+\frac{\dot{A}\dot{B}}{AB}+\frac{\dot{B}\dot{C}}{BC}+\frac{\dot{C}\dot{A}}{CA}].
\end{equation}
With the help of Eq.(\ref{10}), we can write
\begin{eqnarray}\label{512}
-\frac{\ddot{B}}{B}-\frac{\ddot{C}}{C}-\frac{2m^2}{A^2}
+\frac{\dot{A}\dot{B}}{AB}+\frac{\dot{C}\dot{A}}{CA}
+\frac{\dot{A}\dot{F}}{AF}-\frac{\ddot{F}}{F}-\frac{\kappa}{F}(\rho+p)=0,\\\label{513}
-\frac{\ddot{A}}{A}-\frac{\ddot{C}}{C}-\frac{2m^2}{A^2}
+\frac{\dot{A}\dot{B}}{AB}+\frac{\dot{B}\dot{C}}{BC}
+\frac{\dot{B}\dot{F}}{BF}-\frac{\ddot{F}}{F}-\frac{\kappa}{F}(\rho+p)=0,\\\label{514}
-\frac{\ddot{A}}{A}-\frac{\ddot{B}}{B}-\frac{2m^2}{A^2}
+\frac{\dot{B}\dot{C}}{BC}+\frac{\dot{C}\dot{A}}{CA}
+\frac{\dot{C}\dot{F}}{CF}-\frac{\ddot{F}}{F}-\frac{\kappa}{F}(\rho+p)=0.
\end{eqnarray}
The $01$-component can be written by using Eq.(\ref{2}) in the
following form
\begin{equation}\label{0513}
2\frac{\dot{A}}{A}-\frac{\dot{B}}{B}-\frac{\dot{C}}{C}=0.
\end{equation}
We discuss solution of these equations using the same procedure as
for the Bianchi type $I$ solutions.

\subsection{Solution of the Field Equations}

Here we get the same equations Eqs.(\ref{015})-(\ref{017}) as obtained
previously with the difference of the constraint equations (using
Eq.(\ref{0513}))
\begin{equation}\label{0524}
p_1=1,\quad p_2={p_3}^{-1}=P,\quad q_1=0,\quad q_2=-q_3=Q.
\end{equation}
Consequently, the metric functions become
\begin{eqnarray}\label{0519}
A=a,\quad B=aP\exp[{Q\int\frac{dt}{a^3F}}],\quad
C=aP^{-1}\exp[{-Q\int\frac{dt}{a^3F}}].
\end{eqnarray}

\subsection{Model of the Universe when $n\neq0$}

The metric coefficients and the directional Hubble parameters
are the same as given in \cite{1}. Further, we note that the mean
generalized Hubble parameter $H$, the volume
scale factor $V$ and expansion scalar $\theta$ turn out to be the
same as for the Bianchi type $I$ spacetime while the value of
shear scalar $\sigma^2$ is
\begin{equation}\label{0401}
\sigma^2=\frac{Q^2}{k^2(nlt+k_1)^{\frac{2}{n}}}.
\end{equation}
The mean anisotropy parameter for this model becomes
\begin{equation}\label{00401}
A=\frac{2Q^2}{3l^2k^2(nlt+k_1)^{\frac{2}{n}-2}}.
\end{equation}
The energy density and pressure of the universe turn out to be
\begin{equation}\label{00411}
\kappa p=\kappa\rho=-\frac{3kl^2}{(nlt+k_1)^{\frac{2}{n}+2}}
-\frac{Q^2+k^2m^2}{k(nlt+k_1)^{\frac{4}{n}}}.
\end{equation}
The entropy of the universe is given by
\begin{equation}\label{910}
TdS=\frac{1}{\kappa}[6kl^3(n-2)(nlt+k_1)^{\frac{1}{n}-3}-
\frac{2l}{k}(Q^2+k^2m^2)(nlt+k_1)^{-\frac{1}{n}-1}].
\end{equation}
The constraint equation turns out to be same as for
the Bianchi type $I$ spacetime with the condition $Q^2+k^2m^2=0$.
$f(R)$ is given by
\begin{equation}\label{412}
f(R)=\frac{k}{2}(nlt+k_1)^\frac{-2}{n}R+3kl^2(n-2)(nlt+k_1)^\frac{-2n-2}{n},
\end{equation}
where $R=6l^2(n-2)(nlt+k_1)^{-2}+ 8m^2(nlt+k_1)^{\frac{-2}{n}}]$.
For a special case, $n=\frac{1}{2}$, it follows that
\begin{eqnarray}\nonumber
f(R)&=&\frac{k}{2}[\frac{-9l^2\pm \sqrt{81l^4+32m^2R}}{2R}]^{-2}R
\\\nonumber&-& \frac{9kl^2}{2}[\frac{-9l^2\pm
\sqrt{81l^4+32m^2}R}{2R}]^{-3}
\end{eqnarray}
which gives $f(R)$ in terms of $R$ only.

\subsection{Model of the Universe when $n=0$.}

Here the metric coefficients, the mean generalized Hubble parameter,
volume scale factor and expansion scalar are the same as given for
Bianchi type $I$ spacetime. However, the value of shear scalar
$\sigma^2$ is
\begin{equation}\label{0463}
\sigma^2=\frac{Q^2}{k^2{k_2}^2}\exp(-2lt),
\end{equation}
and the mean anisotropy parameter $A$ here becomes
\begin{equation}\label{00463}
A=\frac{2Q^2\exp(-2lt)}{3l^2k^2{k_2}^2}.
\end{equation}
The energy density and pressure of the universe
are
\begin{equation}\label{000463}
\kappa p=\kappa\rho=-\frac{3kl^2\exp(-2lt)}{{k_2}^2}
-(\frac{(Q^2+m^2k^2)}{k{k_2}^4})\exp(-4lt).
\end{equation}
The entropy is
\begin{equation}\label{914}
TdS=\frac{1}{\kappa}[-12kk_2l^3\exp(lt)-
\frac{2l}{kk_2}(Q^2+m^2k^2)\exp(-lt)].
\end{equation}
The constraint equation with the condition,
$Q^2+m^2k^2=0$, turns out to be same as given by Eq.(\ref{0146}).
The function of Ricci scalar, $f(R)$, takes the form
\begin{equation}\label{477}
f(R)=\frac{k}{{2k_2}^2}\exp(-2lt)(R-12l^2)
\end{equation}
with $R=-12l^2+(\frac{8m^2}{{k_2}^2})\exp(-2lt)$. Here we can get
$f(R)$ in terms of $R$
\begin{equation}\label{0477}
f(R)=\frac{k}{16m^2}[R^2-144l^4]
\end{equation}
which also corresponds to the general function $f(R)$ given by
Eq.(\ref{0047}).

\section{Summary and Conclusion}

This paper is to devoted to study the universe expansion in metric
$f(R)$ gravity. We have found exact solutions of the Bianchi types
$I$ and $V$ spacetimes using the non-vacuum field equations. These
solutions correspond to two models of the universe, i.e., a singular
model and a non-singular model. We have evaluated some important
cosmological physical quantities for these solutions such as
expansion scalar $\theta$, shear scalar $\sigma^2$ and mean
anisotropy parameter $A$. The entropy of the universe is also found.
We would like to mention here that solutions for both the spacetimes
correspond to perfect fluid \cite{020, 021} in GR. It is found that
the general function $f(R)$ includes squared power of the Ricci
scalar for the non-singular model.

Firstly we investigate singular model ($n\neq0$) of the universe
which has singularity at $t\equiv t_s=-\frac{k_1}{nl}$. The physical
parameters $H_1,~H_2,~H_3,~H,~\theta$, and $\sigma^2$ are all
infinite at this point for  $n>0$ but the volume scale factor
vanishes. The mean anisotropy parameter $A$ is also infinite at this
point for $0<n<1$ and it will vanish for $n>1$. The function of the
Ricci scalar, energy density, pressure and $T$ are also infinite
while the metric functions vanish at this point of singularity. The
model also suggests that the expansion and shear scalar decrease for
$n>0$ with the passage of time. The mean anisotropy parameter also
decreases for $n>1$ with the increase in time. This indicates that
after a large time the expansion will stop completely and the
universe will achieve isotropy. The isotropy condition, i.e.,
$\frac{\sigma^2}{\theta}\rightarrow 0$ as $t\rightarrow \infty$, is
also satisfied. The entropy of the universe is infinite for
$n>\frac{1}{3}$. Moreover, the second law of thermodynamics is valid
for $n>2$ with the conditions ${q_1}^2+{q_2}^2+{q_3}^2=0$ and
$Q^2+k^2m^2=0$. Thus we can conclude from these observations that
the model starts its expansion from zero volume with infinite energy
density and pressure at $t=t_s$ and it continues to expand with
time.

The second model is non-singular model ($n=0$) of the universe. The
physical parameters $H_1,~H_2,~H_3,~\sigma^2$ and $A$ are all finite
for all finite values of $t$. The mean generalized Hubble parameter
$H$ and expansion scalar $~\theta$ is constant while $f(R)$ is also
finite here. The metric functions do not vanish for this model. The
entropy of the universe is finite while the second law of
thermodynamics is valid for $kk_2<0$ with the conditions
${q_1}^2+{q_2}^2+{q_3}^2=0$ and $Q^2+k^2m^2=0$. The energy density
and pressure become infinite as $t\rightarrow -\infty$ which shows
that the universe started its evolution in an infinite past with a
strong pressure and energy density. The isotropy condition is also
verified for this model. The volume scale factor increases
exponentially with time which indicates that the universe starts its
expansion with zero volume from infinite past.

We would like to mention here that qualitative analysis of the
solutions found has not been done. However, it would be interesting
to perform simulation procedure for the consistency of the results
with Wilkinson Microwave Anisotropy Probe (WMAP) data as given in
the literature.


\vspace{0.05cm}

\begin{thebibliography}{40}

\bibitem{1}Sharif, M. and Shamir, M.F.: Class. Quantum Grav. \textbf{26}(2009)235020.

\bibitem{2}Carmeli, M.: Commun. Theor. Phys.
\textbf{5}(1996)159; Riess, A.G. et al. (Supernova Search Team):
Astron. J. \textbf{116}(1998)1009; Bennett., C.L. et al.:
Astrophys. J. Suppl. \textbf{148}(2003)1; Riess, A.G. et al.:
Astrophys. J. \textbf{607}(2004)665; Spergel, D.N. et al.:
Astrophys. J. Suppl. \textbf{170}(2007)377.

\bibitem{3}Carroll, S.M.: Living Rev. Relativity
\textbf{4}(2001)1.

\bibitem{4}Nojiri, S. and Odintsov, S.D.: Int. J. Geom. Meth. Mod. Phys.
\textbf{115}(2007)4.

\bibitem{05}Lobo, F.S.N.: \emph{The Dark Side of Gravity: Modified Theories of Gravity},
invited chapter to appear in an edited collection "Dark
Energy-Current Advances and Ideas"; arXiv:0807.1640.


\bibitem{005}Faraoni, V.: \emph{f(R) Gravity: Successes and Challenges},
Presented at SIGRAV 2008, 18th Congress of the Italian Society of
General Relativity and Gravitation, Cosenza, Italy September 22-25,
2008; arXiv:0810.2602.

\bibitem{0005}Sotiriou, T.P. and Faraoni, V.: Rev. Mod. Phys. \textbf{82}(2010)451.

\bibitem{5}Nojiri, S. and Odintsov, S.D.: \emph{Problems of Modern Theoretical Physics},
A Volume in honour of Prof. Buchbinder, I.L. On the occasion of his
60th birthday, p.266-285, (TSPU publishing, Tomsk); arXiv:0807.0685.

\bibitem{6}Multam$\ddot{a}$ki, T. and Vilja, I.: Phys. Rev. \textbf{D74}(2006)064022.

\bibitem{7}Multam$\ddot{a}$ki, T. and Vilja, I.: Phys. Rev. \textbf{D76}(2007)064021.

\bibitem{8}Cognola, G., Elizalde, E., Nojiri, S., Odintsov, S.D. and Zerbini, S.: JCAP
\textbf{0502}(2005)010.

\bibitem{9}Capozziello, S., Stabile, A. and Troisi, A.: Class. Quantum Grav.
\textbf{24}(2007)2153.

\bibitem{010}Sharif, M. and Kausar, H.R.: \textit{Non-vacuum Static Spherically
Symmetric Solutions in $f(R)$ Gravity}, submitted for publication.

\bibitem{10}Hollenstein, L. and Lobo, F.S.N.: Phys. Rev.
\textbf{D78}(2008)124007.

\bibitem{11}Azadi, A., Momeni, D. and Nouri-Zonoz, M.: Phys. Lett.
\textbf{B670}(2008)210.

\bibitem{12}Momeni, D. and Gholizade, H.: Int. J. Mod. Phys.
\textbf{D18}(2009)1.

\bibitem{13}Sharif, M. and Shamir, M.F.: Mod. Phys. Lett. \textbf{A25}(2010)1281.

\bibitem{0018}Lorenz-Petzold, D.: Phys. Rev.
\textbf{D29}(1984)2399.

\bibitem{0019}Lorenz-Petzold, D.: Astrophys. Space Sci.
\textbf{114}(1985)277.

\bibitem{019}Berman, M.S.: Nuovo Cimento
\textbf{B74}(1983)182.

\bibitem{022}Kumar, S. and Singh, C.P.: Int. J.
Theor. Phys. \textbf{47}(2008)1722.

\bibitem{0220}Singh, C.P.: Brazilian J. Phys.
\textbf{39}(2009)4.

\bibitem{0022}Paul, B.C., Debnath, P.S. and Ghose, S.: Phys. Rev.
\textbf{D79}(2009)083534.

\bibitem{16}Nojiri, S. and Odintsov, S.D.: Phys. Rev.
\textbf{D68}(2003)123512.

\bibitem{020}Singh, C.P., Ram, S. and Zeyauddin, M.: Astrophys. Space Sci.
\textbf{315}(2008)181.

\bibitem{021}Kumar, S. and Singh, C.P.: Astrophys. Space
Sci. \textbf{312}(2007)57.

\end{thebibliography}
\end{document}